\documentclass[aps, pra, superscriptaddress, twocolumn, amsfonts, amsmath, amssymb, floatfix]{revtex4-2}
\usepackage{graphicx}
\usepackage{float}
\usepackage{sidecap}
\usepackage{dcolumn}
\usepackage{bm}
\usepackage{epsfig}
\usepackage{braket}
\usepackage{color}
\usepackage{ulem}
\usepackage{amsmath}
\usepackage[colorlinks=true, letterpaper=true, pdfstartview=FitV, linkcolor=blue, citecolor=blue, urlcolor=blue]{hyperref}

\begin{document}

\title{Scissors Modes of a Bose-Einstein Condensate in a Synthetic Magnetic Field}

\author{Chunlei Qu}
\email{Corresponding author. cqu5@stevens.edu}
\affiliation{Department of Physics, Stevens Institute of Technology, Hoboken, NJ 07030, USA}
\affiliation{Center for Quantum Science and Engineering, Stevens Institute of Technology, Hoboken, NJ 07030, USA}

\author{Chuan-Hsun Li}
\affiliation{Department of Physics and Astronomy, Purdue University, West Lafayette, Indiana 47907, USA}

\author{Yong P. Chen}
\affiliation{Department of Physics and Astronomy, Purdue University, West Lafayette, Indiana 47907, USA}
\affiliation{School of Electrical and Computer Engineering, Purdue University, West Lafayette, Indiana 47907, USA}
\affiliation{Purdue Quantum Science and Engineering Institute, Purdue University, West Lafayette, Indiana 47907, USA}

\author{Sandro Stringari}
\affiliation{Pitaevskii BEC Center and Dipartimento di Fisica, Universit\`a degli Studi di Trento, 38123 Povo, Italy}

\date{\today}
\begin{abstract}
We study the scissors modes of a harmonically trapped Bose-Einstein condensate under the influence of a synthetic magnetic field, which induces rigid rotational components in the velocity field. Our investigation reveals that the scissors mode, excited in the plane perpendicular to the synthetic magnetic field, becomes coupled to the quadrupole modes of the condensate, giving rise to typical beating effects. Moreover, the two scissors modes excited in the vertical planes are also coupled together by the synthetic magnetic field, resulting in intriguing gyroscope dynamics. Our analytical results, derived from a spinor hydrodynamic theory, are further validated through numerical simulations of the three-dimensional Gross-Pitaevskii equation. These predictions for the condensates subject to a synthetic magnetic field are experimentally accessible with current cold-atom setups and hold promise for potential applications in quantum sensing. 
\end{abstract}

\maketitle

\textcolor{blue}{\textit{Introduction.---}}
Superfluidity is one of the most extraordinary consequences of Bose-Einstein condensation (BEC). In usual condensates, superfluidity is manifested in various remarkable rotational phenomena such as a reduced moment of inertia relative to the rigid-body value and the vanishing of angular momentum in absence of quantized vortices and in presence of isotropic trapping~\cite{PhysRevLett.76.1405,BECbook}. Both effects are directly linked to the constraint of irrotationality, a peculiar feature of usual superfluids.  In vortex-free BECs, superfluidity is often probed by exciting the so-called scissors mode which corresponds to an angular rotation around the symmetry axis of an anisotropic trap~\cite{PhysRevLett.83.4452}. This technique has proved invaluable in discerning the superfluid nature of diverse systems, including Bose~\cite{PhysRevLett.84.2056} and Fermi gases~\cite{PhysRevLett.99.150403} as well as elongated dipolar atomic gases~\cite{PhysRevLett.120.160402,doi:10.1126/science.aba4309}.

\begin{figure}[t!]
\centerline{
\includegraphics[width=0.5\textwidth]{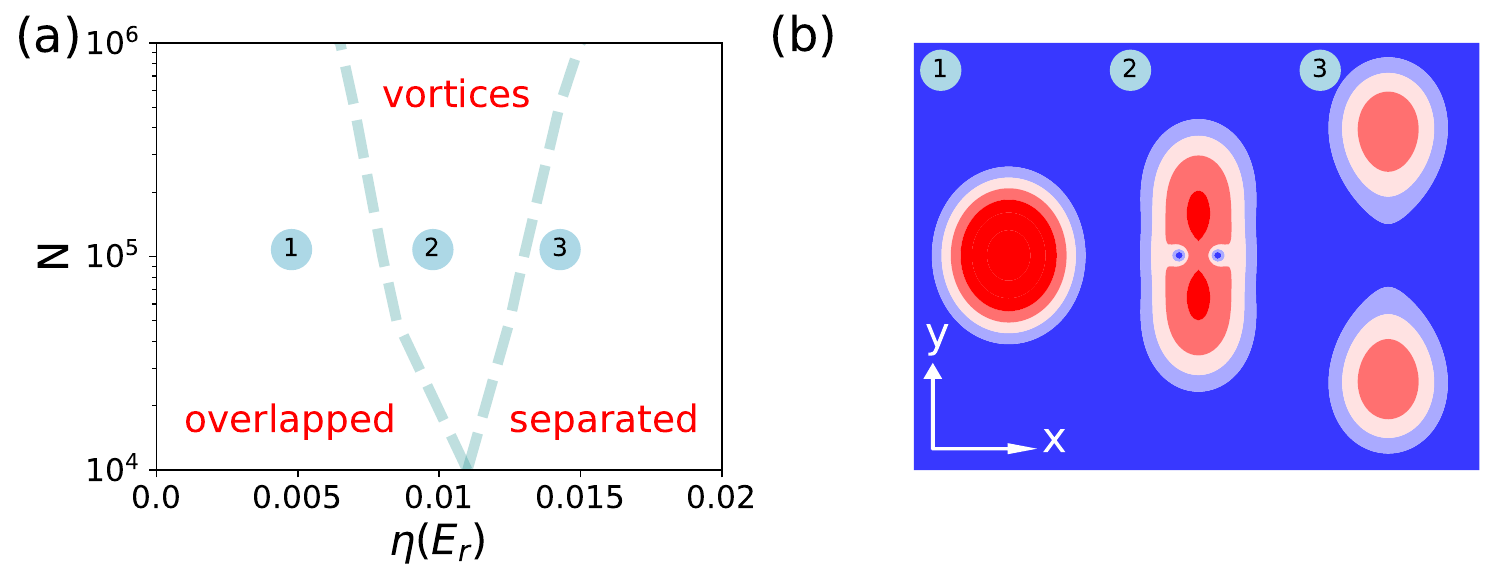}}
\caption{(a) Phase diagram of the condensate in a synthetic magnetic field. The two spin components can be overlapped, exhibit vortices, or be separated, depending on the value of the relevant parameter $\eta$ fixing the position dependence of the detuning (see text) and on the particle number $N$. (b) A typical distribution of the condensate for $\eta=0.005, 0.01, 0.013$ at $N=10^5$. Note that there are two vortices at $y=0$ for $\eta=0.01$.}
\label{fig:vortex}
\end{figure}

The experimental realization of synthetic spin-orbit coupling (SOC) has offered novel opportunities for exploring topological condensed matter physics with ultracold neutral atoms~\cite{RevModPhys.83.1523,galitski2013spin,RevModPhys.91.015005}. Synthetic SOC substantially impacts the superfluid properties of condensates due to the breaking of Galilean invariance and irrotationality~\cite{lin2009synthetic,PhysRevLett.105.160403,PhysRevLett.107.150403,wu2011unconventional,PhysRevA.84.063604,zhu2012exotic,PhysRevLett.108.225301,PhysRevLett.108.010402,PhysRevLett.108.035302,PhysRevA.87.063610,PhysRevA.94.033635,PhysRevLett.118.145302,PhysRevLett.120.183202}. For instance, in BECs with an equal superposition of Rashba and Dresselhaus SOC, the superfluid fraction associated with the flow along the SOC direction vanishes near the transition point between the plane-wave phase and the zero-momentum phase~\cite{PhysRevA.94.033635}. Furthermore, introducing a position-dependent detuning along the direction orthogonal to the SOC direction generates new intriguing features in the phase diagram~\cite{PhysRevA.84.063604}. Figure~\ref{fig:vortex} shows that, as a function of the position-dependent detuning, the two spin components of spin-orbit-coupled BEC can overlap, exhibit vortices, or be fully separated. Vortices have been actually observed experimentally for values of the detuning gradient larger than a critical value~\cite{lin2009synthetic}. Despite these advances, the effect of the synthetic magnetic field on the superfluid dynamics of the spin-orbit-coupled condensates has been vastly unexplored. Recently it was found that a vortex-free BEC, in the presence of position-dependent detuning, exhibits a rigid-like rotational velocity field and possesses a finite angular momentum even when the system is confined in an isotropic harmonic trap~\cite{PhysRevLett.118.145302}. Due to spin-orbit coupling the position-dependent detuning actually brings the system into an effective non-inertial rotational frame, leading to the Foucault-like precession of two dipole modes in the rotating plane~\cite{PhysRevLett.120.183202}.

In this paper, we extend the spinor hydrodynamic theory developed in~\cite{PhysRevLett.118.145302} and \cite{PhysRevLett.120.183202}, accounting for position-dependent tuning effects, in order to investigate the precession of the scissors modes of a BEC subject to a rotational velocity field induced by a synthetic magnetic field. Our findings indicate that the synthetic magnetic field induces a coupling between the scissors modes and the other collective modes of quadrupole nature. Specifically, the scissors mode in the horizontal plane perpendicular to the synthetic magnetic field becomes coupled to the three quadrupole modes related to the shape oscillation of the condensate. Moreover, the two scissors modes in the vertical planes are coupled together, giving rise to intriguing gyroscope dynamics.

\begin{figure}[t!]
\centerline{
\includegraphics[width=0.5\textwidth]{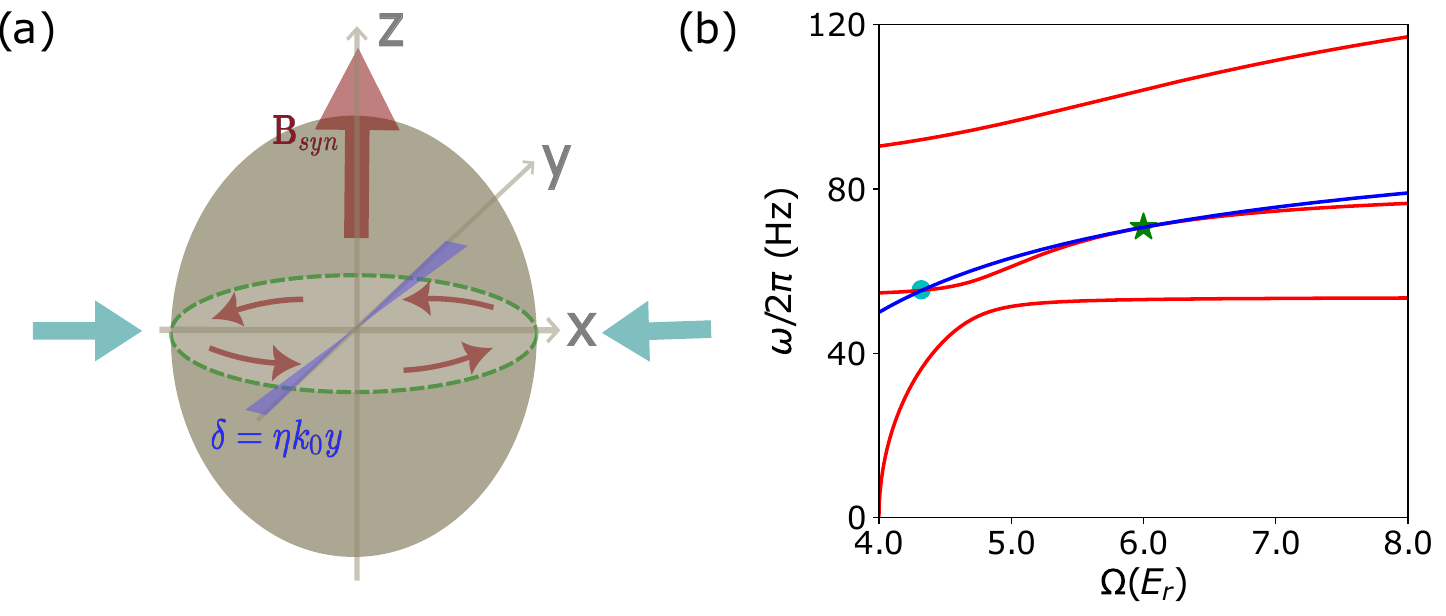}}
\caption{(a) Illustration of a spin-orbit-coupled BEC in the presence of a position-dependent detuning $\delta(y) = \eta k_0 y$ which induces a synthetic magnetic field $\mathbf{B}_{syn}$ along the $z$-direction and a rotational velocity field in the horizontal plane. (b) Dependence of the quadrupole modes eigenfrequencies on the Raman coupling strength for $\eta=0$. The three red curves are the eigenfrequencies corresponding to the width oscillation of the condensate (Eq.~\ref{eq:quad}) and the blue curve corresponds to the  oscillation frequency $\omega_{xy}/2\pi$ of the scissors mode in the horizontal plane. The trapping frequencies are $(\omega_x, \omega_y, \omega_z)/2\pi=(50\sqrt{3}, 50, 35)$Hz. At the two points indicated by the green star and cyan dot, the scissors mode frequency $\omega_{xy}$ and one of the three hybrid quadrupole modes are degenerate. In this work, we shall focus on the green star point at $\Omega=6E_r$ where the degeneracy condition Eq.~(\ref{eq:cond1}) is satisfied.}
\label{fig:illustration}
\end{figure}

\textcolor{blue}{\textit{Spinor hydrodynamic theory.---}}We begin by considering a BEC with a Raman transition induced equal-Rashba-Dresselhaus synthetic SOC along the $x$-direction. In the spin-rotated frame, the single-particle Hamiltonian is given by 
\begin{equation}
H_{0}=\frac{1}{2m}(\hat{\mathbf{p}}-\hbar k_0 \sigma_z \hat{e}_x )^2+ V_{trap} -\frac{\hbar\Omega}{2}\sigma_x
-\eta k_0y\sigma_z,
\label{eq:sp}
\end{equation}
where $m$ is the atomic mass, $\hbar k_0$ is the recoil momentum, $\Omega$ represents the Raman coupling strength, $V_{trap}=m(\omega_x^2 x^2 + \omega_y^2 y^2 + \omega_z^2 z^2)/2$ denotes the harmonic trapping potential with angular frequencies $\omega_{x,y,z}$, and $\eta$ is the coefficient of the $y$-position-dependent detuning. The mean-field interaction energy of the condensate is characterized by $V_{int}=\int d\mathbf{r} \sum_{\alpha\beta}g_{\alpha\beta}n_\alpha n_\beta/2$ where $n_\alpha$ is the particle density of the $\alpha$-th component, and $g_{\alpha\beta}=4\pi\hbar^2 a_{\alpha\beta}/m$ are the interaction constants in different spin channels with $a_{\alpha\beta}$ the corresponding $s$-wave scattering lengths. In this work, we focus on isotropic interactions by assuming $g_{\alpha\beta}=g$. The last term in Eq.~(\ref{eq:sp}) induces a synthetic magnetic field $\mathbf{B}_{syn}$ along the vertical $z$-direction (Fig.~\ref{fig:illustration}(a)). When $\eta$ is larger than a critical value $\eta_{c}$, vortices will appear which has been demonstrated in the pioneering experiment~\cite{lin2009synthetic}. 

For simplicity, we shall focus on the zero-momentum phase where $\hbar\Omega>\hbar\Omega_c\equiv 4E_r$, with $E_r=\hbar^2k_0^2/2m$ representing the recoil energy. When the detuning gradient $\eta$ is small, the total density of the condensate can be approximated using the Thomas-Fermi (TF) distribution. Additionally, due to the large Raman energy gap, the relative phase between the two spin components is locked in the low-frequency oscillations. In this regime the hydrodynamic equations for the two spin components can be simply reduced to the following  equations
\begin{eqnarray}
&&\frac{\partial n}{\partial t} + \mathbf{\nabla}\cdot (n\mathbf{v}) = 0 \label{eq:continuity} \\
&& 
\frac{\partial\phi}{\partial t} 
+
\frac{1}{2}m^*v_x^2 + \frac{1}{2}mv_y^2 + \frac{1}{2}mv_z^2 + \mu - \frac{\Omega}{2}=0
\label{eq:hd2}
\end{eqnarray}
for the total density $n$, and the the absolute phase $\phi$, with
\begin{equation}
\mathbf{v}=\left(
\frac{\hbar}{m^*}\nabla_x\phi - \frac{\eta}{\hbar}\frac{\Omega_c}{\Omega}y,
\frac{\hbar}{m}\nabla_y\phi,
\frac{\hbar}{m}\nabla_z\phi
\right),
\end{equation}
denoting the physical superfluid velocity and $m^*=m(1-\Omega_c/\Omega)^{-1}$ the effective mass in the zero-momentum phase. It is worth pointing out that the second term of $v_x$ arises from the spin density $s_z/n$ which becomes nonzero for finite $\eta$. This term plays an important role in generating the rotational velocity field, the finite angular momentum, and the coupling among different collective modes. In Eq.~(\ref{eq:hd2}), we have introduced the chemical potential $\mu = ng + V_{trap}-\frac{1}{2}m^*
\eta^2\Omega_c^2y^2/\hbar^2\Omega^2$
where the last term results from the anti-trap effect of the synthetic magnetic field and renormalizes the trapping frequency along the $y$-direction. Since it is proportional to $\eta^2$, this term is neglected in the following analysis.

At equilibrium, the total density and absolute phase are determined as
$n_0=(\mu-V_{trap})/g$ and $\phi_0=\alpha xy$, respectively, with $\alpha=2\eta k_0^2\omega_x^2/\Omega \omega_{xy}^2$ where $\omega_{xy}=\sqrt{\omega_x^2m/m^*+\omega_y^2}$ is the frequency of the scissors mode in the horizontal plane in the absence of detuning gradient. Consequently, the velocity field becomes $\mathbf{v}_0=(-\omega_{eff}y, \omega_{eff}^\prime x, 0)$ where $\omega_{eff}=\eta\Omega_c/\hbar\Omega - \hbar\alpha/m^*$, $\omega_{eff}^\prime=\hbar\alpha/m$. 
Such a rigid-like rotational velocity field causes a coupling of various collective modes, including the two dipole modes in the horizontal plane~\cite{PhysRevLett.120.183202} and the quadrupole modes discussed in this work.

\textcolor{blue}{Without synthetic magnetic field ($\eta=0$).---} In this case the quadrupole modes of a trapped BEC include three scissors modes excited by the operators $xy$, $xz$, and $yz$ and corresponding, respectively,  to the angular rotation of the condensate around the $z$, $y$, and $x$ axis, and other three modes excited by the operators $x^2$, $y^2$, and $z^2$, which correspond to the shape oscillations of the condensate. The oscillation frequencies of the three scissors modes are given by
$\omega_{xy} = \sqrt{\omega_x^2(m/m^*)+\omega_y^2}$, 
$\omega_{xz} = \sqrt{\omega_x^2(m/m^*)+\omega_z^2}$,
$\omega_{yz} = \sqrt{\omega_y^2+\omega_z^2}$, respectively. 

The frequencies of the quadrupole modes related to the shape oscillations are instead given by the solution of the equation~\cite{BECbook} 
\begin{eqnarray}
&&\omega^6-3\omega^4(\frac{m}{m^*}\omega_x^2+\omega_y^2+\omega_z^2)+8\omega^2(\frac{m}{m^*}\omega_x^2\omega_y^2+\frac{m}{m^*}\omega_x^2\omega_z^2 \notag  \\
&&
+\omega_y^2\omega_z^2)-20\frac{m}{m^*}\omega_x^2\omega_y^2\omega_z^2=0.
\label{eq:quad}
\end{eqnarray}
Figure~\ref{fig:illustration}(b) illustrates how the three eigenfrequencies related to the shape oscillations and the scissors mode oscillation frequency $\omega_{xy}$ depend on the Raman coupling strength. It shows that the degeneracy between the scissors mode $xy$ and one of the shape oscillation quadrupole modes takes place for two values of the Raman coupling. In the following we will focus on the case indicated with the green star point at $\Omega = 6E_r$, corresponding to the degeneracy condition~\cite{cond2} (for the more general case, and detailed information on the derivation of the equations of motion for the collective modes, see SM~\cite{SM}) 
\begin{eqnarray}
\sqrt{\frac{m}{m^*}}\omega_x = \omega_y \label{eq:cond1}
\end{eqnarray}
which actually coincides with the condition that the two dipole modes in the horizontal $x-y$ plane be degenerate.  

\textcolor{blue}{With synthetic magnetic field ($\eta\neq 0$).---}In the presence of a position-dependent detuning, the six quadrupole modes discussed above are further coupled together by the synthetic magnetic field, leading to interesting beating phenomena that can be induced through distinct excitation methods:

\begin{figure}[t!]
\centerline{
\includegraphics[width=0.5\textwidth]{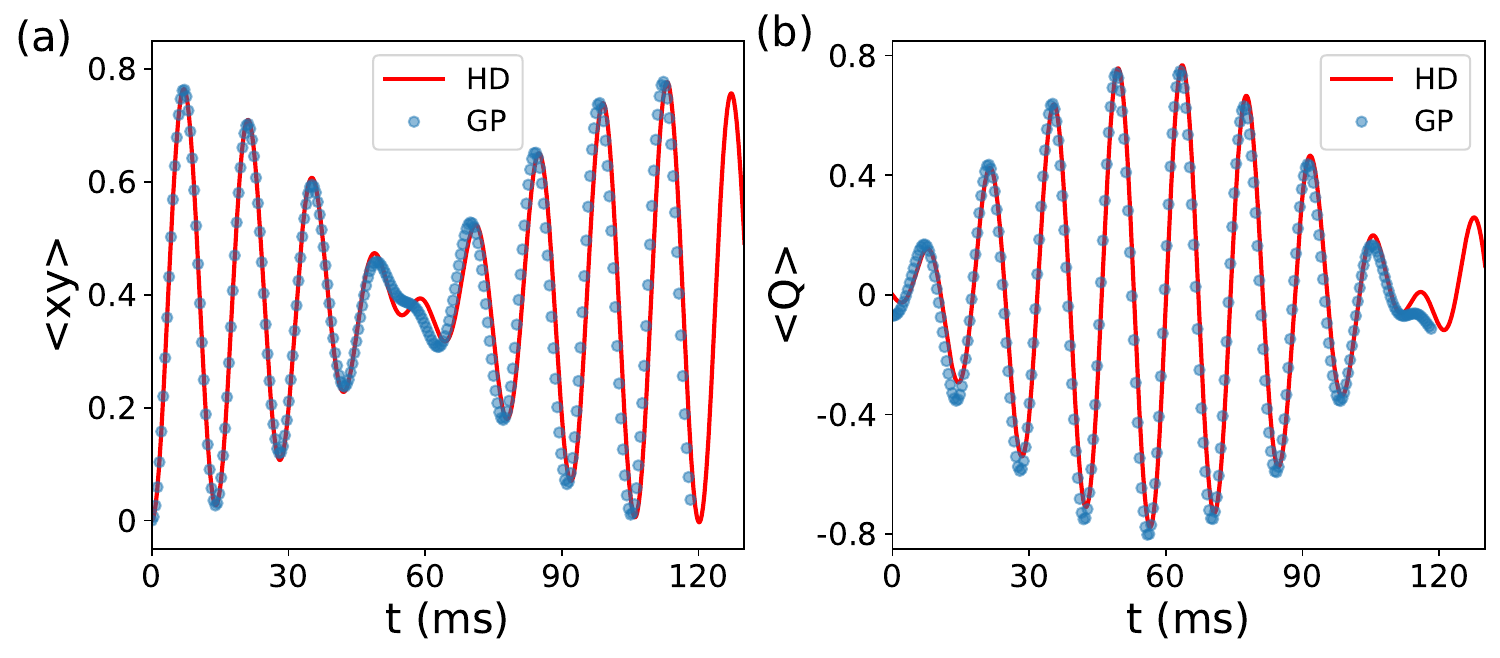}}
\caption{(a,b) Time evolution of the scissors mode $xy$ in the horizontal plane and the quadrupole mode characterized by the operator $Q=(\omega_x^2/\omega_y^2)x^2-y^2$ in the presence of a detuning gradient $\eta=0.001E_r$. The trapping frequencies are $(f_x, f_y, f_z)=(50\sqrt{3}, 50, 35)$Hz and the Raman coupling strength is $\Omega=6E_r$ to meet the degeneracy condition Eq.~(\ref{eq:cond1}). The dynamics is excited by a sudden rotation of the harmonic trap by a small angle $\varphi_0=3^\circ$ in the $x-y$ plane.}
\label{fig:scissors}
\end{figure}

\textcolor{blue}{\textit{i) Sudden rotation of the harmonic trap by a small angle $\varphi_0$ in the x-y plane.}} This excites the scissors mode in the horizontal plane, characterized by the operator $xy$. Due to the influence of the synthetic magnetic field, this scissors mode becomes coupled to the three quadrupole modes $x^2$, $y^2$, and $z^2$, related to the shape oscillations of the condensate. The ansatz for the fluctuation of the total density and absolute phase can be written as
\begin{eqnarray}
\delta n &\sim& \epsilon_1 xy+\epsilon_2 x^2 +\epsilon_3 y^2+\epsilon_4 z^2 \label{eq:dn1}  \\ 
\delta\phi   &\sim& \alpha_1 xy+\alpha_2 x^2+\alpha_3 y^2+\alpha_4 z^2
\label{eq:dphi1}
\end{eqnarray}
Substituting them into the linearized spinor hydrodynamic Eqs.~(\ref{eq:continuity}) and (\ref{eq:hd2}), we obtain the coupled differential equations for the variables $\epsilon_j$ and $\alpha_j$ ($j=1,\ldots,4$) (see SM~\cite{SM}). By seeking solutions of the form $\epsilon_j(t)\sim e^{-i\omega t}$ and $\alpha_j(t)\sim e^{-i\omega t}$, the collective frequencies are found to satisfy an equation of the form $\omega^8+c_3\omega^6+c_2\omega^4+c_1\omega^2+c_0=0$. 

\begin{figure}[t!]
\centerline{
\includegraphics[width=0.5\textwidth]{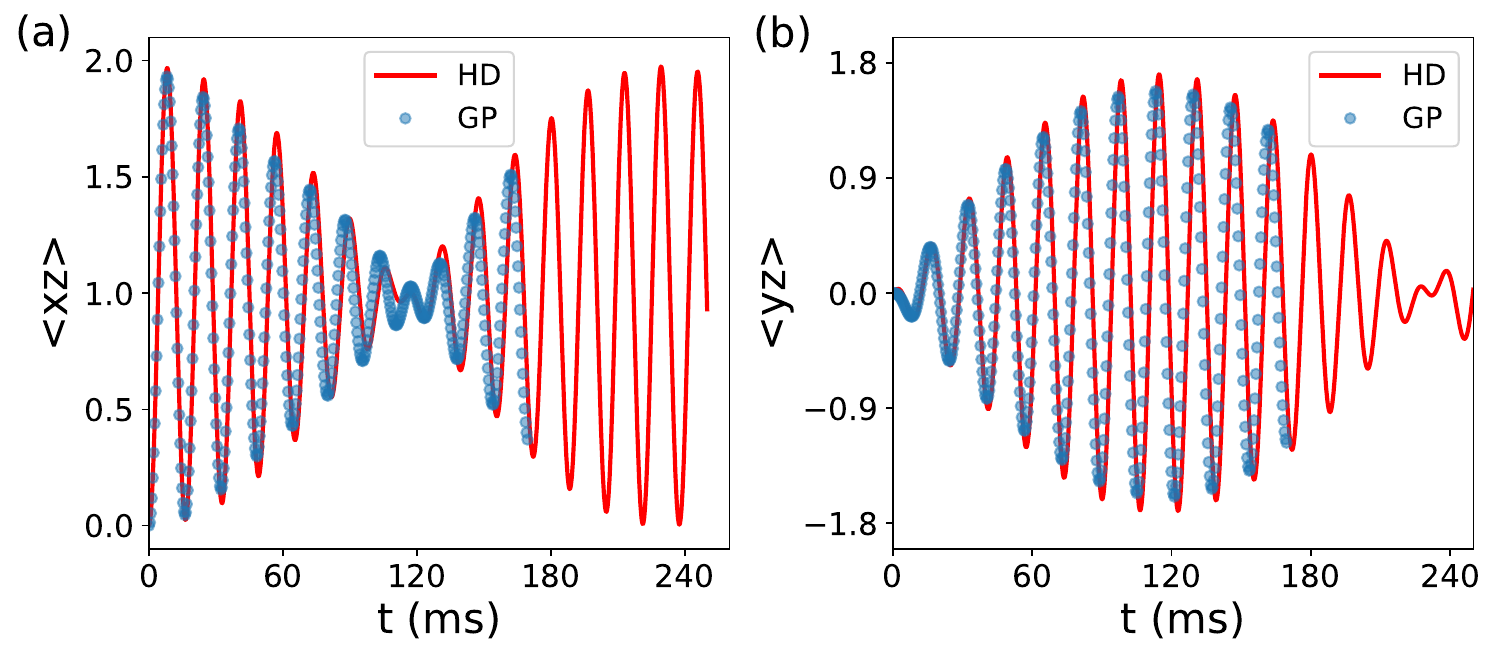}}
\caption{Time evolution of the scissors modes in the vertical planes $\langle xz \rangle$ and $\langle yz\rangle$. The trapping frequencies are $(f_x, f_y, f_z)=(50\sqrt{3}, 50, 35)Hz$, $\Omega=6E_r$. Thus, $\sqrt{m/m^*}\omega_x=\omega_y$, a beat appears for the two coupled modes. The detuning gradient $\eta=0.001E_r$. The dynamics is excited by a sudden rotation of the harmonic trap by a small angle $\theta_0=3^\circ$ in the $x-z$ plane.}
\label{fig:gyroscope}
\end{figure}

Under the degeneracy condition (Eq.~(\ref{eq:cond1})), these coupled differential equations can be further simplified as the scissors mode $xy$ is only coupled to a mode associated with condensate deformation in the horizontal plane, excited by the operator $Q=\frac{\omega_x}{\omega_y}x^2-\frac{\omega_y}{\omega_x} y^2$. Introducing the convenient variables $\sqrt{\frac{m}{m^*}}\omega_x = \omega_y\equiv\omega_0$ and $\omega_{eff}\frac{\omega_x}{\omega_y} = \omega_{eff}^\prime \frac{\omega_y}{\omega_x}\equiv \omega_{\eta}$, a straightforward calculation, which includes only effects linear in $\eta$, yields (see SM~\cite{SM})
\begin{eqnarray}
&&\frac{d^2}{dt^2}\langle xy \rangle + (2\omega_0^2-4\omega_\eta^2)\langle xy \rangle  - 2\omega_\eta \frac{d}{dt}\langle Q \rangle = 0 \\ 
&& \frac{d^2}{dt^2}\langle Q \rangle + (2\omega_0^2-4\omega_\eta^2)\langle Q \rangle + 8\omega_\eta \frac{d}{dt} \langle xy \rangle = 0.
\end{eqnarray}
Consequently, the two modes excited by the operators $xy$ and $Q$ exhibit a beating effect with a frequency splitting $\Delta\omega=4\omega_{\eta}$. The other two modes, excited by the operators $\frac{\omega_x}{\omega_y}x^2+\frac{\omega_y}{\omega_x} y^2$ and $z^2$, remain coupled but they do not depend on the detuning gradient (see SM~\cite{SM}). A representative numerical solution of the beating effect exhibited by the $xy$ and $Q$ modes is shown in Fig.~\ref{fig:scissors}(a, b). 

To validate our theoretical hydrodynamic analysis, we have performed a numerical simulation of the time-dependent three-dimensional Gross-Pitaevskii (GP) equation. For our numerics, we have chosen a total particle number of $N=10^5$, and a recoil momentum $k_0=2\pi/\lambda$ with $\lambda=782nm$. Our results reveal excellent agreement between the hydrodynamic predictions and numerical GP results. 

\textcolor{blue}{\textit{ii) Sudden rotation of the harmonic trap by a small angle $\theta_0$ in the x-z plane.}} This excites the scissors mode in the vertical plane, associated with the operator $xz$. Due to the presence of the synthetic magnetic field, this mode couples to the other scissors mode $yz$. In this case the ansatz for the variations of the total density and absolute phase are
\begin{eqnarray}
&&\delta n \sim \epsilon_5 xz+\epsilon_6 yz \label{eq:dn2}
\\
&&\delta \phi \sim \alpha_5 xz+\alpha_6 yz
\label{eq:dphi2}
\end{eqnarray}
Under the same degeneracy condition Eq.~(\ref{eq:cond1}), the two scissors modes share the same oscillation frequency when $\eta=0$. Substituting Eqs.~(\ref{eq:dn2}-\ref{eq:dphi2}) into the linearized spinor hydrodynamic Eqs.~(\ref{eq:continuity}-\ref{eq:hd2}), we obtain the following coupled second-order differential equations 
\begin{eqnarray}
\frac{d^2\langle xz\rangle}{dt^2} +(\omega_0^2+\omega_z^2-\omega_\eta^2)\langle xz \rangle +2\omega_{\eta} \frac{\omega_y}{\omega_x}\frac{d\langle yz \rangle}{dt} = 0 \notag \\
\frac{d^2\langle yz \rangle}{dt^2} +(\omega_0^2+\omega_z^2-\omega_\eta^2)\langle yz \rangle  -2\omega_{\eta}
\frac{\omega_x}{\omega_y}\frac{d\langle xz \rangle }{dt} = 0 \notag
\end{eqnarray}
which clearly reveals that the two scissors modes exhibit a beating effect, with frequency splitting $\Delta\omega=2\omega_\eta$ (Fig.~\ref{fig:gyroscope}).

\begin{figure}[t!]
\centerline{
\includegraphics[width=0.18\textwidth]{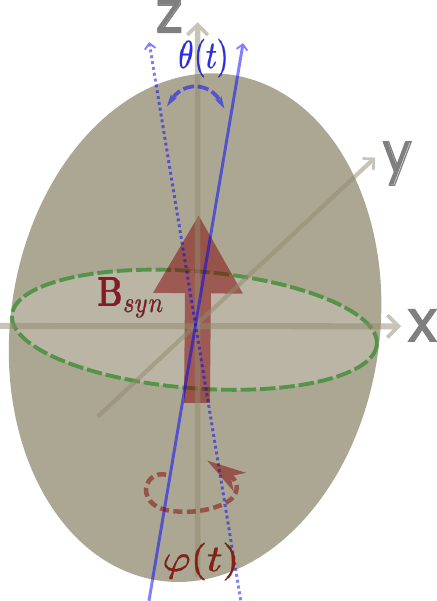}}
\caption{Illustration of a BEC gyroscope realized via a synthetic magnetic field. The polar angle $\theta(t)$ corresponds to a fast scissors mode oscillation in the vertical plane and the azimuthal angle $\varphi(t)$ corresponds to a slower precession around the symmetry axis $z$ due to the synthetic magnetic field.}
\label{fig:gyroscope2}
\end{figure}

The coupling between the two scissors modes in the vertical planes gives rise to a typical gyroscopic effect associated with the precession of angular momentum. A sudden rotation of the trap in the $x-z$ plane in fact causes the angular momentum to deviate from the vertical axis. As shown in Fig.~\ref{fig:gyroscope2}, this results in two simultaneous dynamical effects: a rapid oscillation of angular momentum around the vertical axis due to the restoring force of the trapping potential (the scissors mode), and a slower precession of angular momentum around the vertical axis due to the synthetic magnetic field.

In the linear regime, it is easy to find the relation between the azimuthal angle $\varphi$, that characterizes the gyroscope precession and the polar angle $\theta$, that characterizes the  scissors modes in the vertical planes:
\begin{eqnarray}
 \langle xz \rangle &=& \frac{R_xR_z}{7}\frac{\omega_x^2-\omega_z^2}{\omega_x\omega_z}\theta(t)\cos\varphi(t) \\
  \langle yz \rangle &=& \frac{R_yR_z}{7}\frac{\omega_y^2-\omega_z^2}{\omega_y\omega_z}\theta(t)\sin\varphi(t)
\end{eqnarray}
where we have assumed that the polar angle $\theta(t)$ is small. A similar gyroscopic effect has been previously investigated in the case of a single component Bose-Einstein condensate in the presence of a quantized vortex~\cite{PhysRevLett.86.4725, PhysRevLett.91.090403}. In the present case the effect is caused by the synthetic magnetic field which gives rise to angular momentum even in the absence of quantized vortices, and to a precession rate $d\varphi/dt$ of the azimuthal angle proportional to $\omega_\eta$.

For larger values of the position-dependent detuning gradient $\eta$, and hence larger synthetic magnetic fields, additional effects take place, like bifurcation dynamics~\cite{PhysRevLett.86.377} and the occurrence of quantized vortices~\cite{lin2009synthetic}, whose consequences on the dynamic behavior of the scissors mode, are however beyond the purpose of the present work.

In conclusion, we have presented a comprehensive investigation of the scissors modes and other quadrupole modes of a spin-orbit-coupled BEC under the influence of a position-dependent detuning. Due to the existence of a synthetic magnetic field, these collective modes are coupled together, giving rise to interesting beating effects and gyroscope dynamics. Our framework can be naturally extended to the case of the plane-wave phase or the application of a position-dependent Raman coupling. Our results offer valuable insights pertinent to the understanding of a recent experimental work carried out at Purdue University where the scissors mode of a spin-orbit-coupled BEC with a position-dependent Raman coupling was directly observed~\cite{Li2022}.

\begin{acknowledgments}
C.Q is supported by the ACC-New Jersey under Contract No. W15QKN-18-D-0040 and the Stevens Startup Fund. C.H.L and Y.P.C acknowledge support from Quantum Science Center (QSC), a Department of Energy (DOE) quantum information science (QIS) research center.
\end{acknowledgments}

\nocite{*}
\bibliography{references}

\begin{thebibliography}{29}
\bibitem [{con()}]{cond2}%
  \BibitemOpen
  \bibfield  {title} {\bibinfo {title} {The other degeneracy condition can be
  analytically obtained as $\omega_z=\sqrt{2(m\omega_x^2/m^*+\omega_y^2)/5}$},\
  }\href@noop {} {\ }\BibitemShut {NoStop}%
\end{thebibliography}

\end{document}